\documentclass[aps,twocolumn,showpacs,floatfix]{revtex4}
\usepackage{graphicx}

\newcommand{\beq}{\begin{equation}}
\newcommand{\eeq}{\end{equation}}
\newcommand{\qsqz}{\frac{q}{q_0}}
\newcommand{\PRL}{Phys. Rev. Lett.~}
\newcommand{\PRA}{Phys. Rev. A~}

\begin{document}

\title{Unstable regimes for a Bose-Einstein condensate in an optical lattice}

\author{L. De~Sarlo$^{*}$, L. Fallani, J. E. Lye, M. Modugno$^{1}$,
R. Saers$^{\dagger}$, C.~Fort  and M. Inguscio}

\affiliation{LENS, Dipartimento di Fisica, and INFM Universit\`a
di Firenze via Nello
Carrara 1, I-50019 Sesto Fiorentino (FI), Italy\\
$^{1}$ also BEC-INFM Center, Universit\`a di Trento, I-38050 Povo
(TN), Italy}

\begin{abstract}
We report on the experimental characterization of energetic and
dynamical instability, two mechanisms responsible for the breakdown
of Bloch waves in a Bose-Einstein condensate interacting with a 1D
optical lattice. A clear separation of these two regimes is obtained
performing measurements at different temperatures of the atomic
sample. The timescales of the two processes have been determined by
measuring the losses induced in the condensate. A simple
phenomenological model is introduced for energetic instability while
a full comparison is made between the experiment and the 3D
Gross-Pitaevskii theory that accounts for dynamical instability.
\end{abstract}


\date{\today}

\maketitle

\section{Introduction}
The interest in the system made by neutral atoms in optical lattices
has constantly been growing in the last decades since the
development of efficient laser cooling techniques, which opened the
possibility of observing quantum effects on the motion of atomic
ensembles. The physics of quantum particles in periodic potentials
can be described in terms of Bloch waves \cite{solidstate} and
indeed many effects originally predicted for electrons in a lattice
of ions have been observed for ultracold thermal atoms moving in
optical lattices \cite{blochosc} and, more recently, in quantum
degenerate samples with the observation of long lived Bloch
oscillations in a degenerate Fermi gas \cite{modugno}. In
particular, the achievement of Bose-Einstein condensation (BEC) in
dilute atomic gases has allowed the possibility to repeat these
experiments with ensembles of particles all occupying the same Bloch
state, in principle enhancing the visibility of these quantum
effects \cite{rolston,lensing,dispman,arimondo}. However, when the
density of the sample increases, as in the case of a trapped
condensate, interactions among the atoms forming the BEC may
significantly change the simple single-particle picture. The
interaction-induced nonlinearity is responsible for the observation
of many other phenomena, notably the phase transition from a
superfluid to a Mott-insulator \cite{MI} and the generation of
bright gap solitons \cite{soliton}. Furthermore, it can be shown
that there exists a range of parameters for which non-linearity
makes the Bloch-like solutions of the wave equation describing the
system unstable.

In this paper we report on the experimental characterization of
the unstable regimes for a BEC in a 1D optical lattice, obtained
through a clear measurement of the timescales describing the
evolution of the system, in remarkable agreement with the theory.

\begin{figure}[ht]
\centering
\includegraphics[width=0.9\columnwidth,clip]{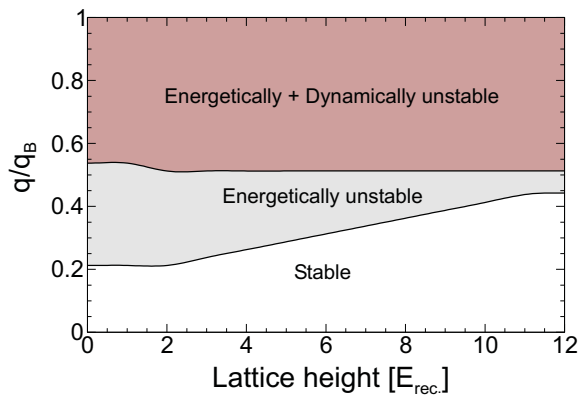}
\caption{Schematic plot of the stability of Bloch matter waves in
the first band of a 1D optical lattice: white area corresponds to
the stable region, in the light gray area the condensate is
energetically unstable and in the dark gray area it is both
energetically and dynamically unstable \cite{wu,mmod}. Vertical axis
is the Bloch quasimomentum $q$ and horizontal axis is the lattice
height in recoil energy (see Sec.~\ref{exp}).}\label{fig0}
\end{figure}

The phenomenon of instability of a BEC in a 1D optical lattice has
already been the subject of experimental \cite{burger,cataliottinjp,
pisani,di} and theoretical
\cite{wu,smerzi,machholm,konotop,scott,zheng,nesi,mmod,carr} works.
In particular, in the framework of the Gross-Pitaevskii (GP) theory
in a periodic potential, there are different mechanisms responsible
for the breakdown of the initial superfluid state. On one hand, as
demonstrated by Landau in the context of superfluid helium
\cite{landaulif}, there is a critical velocity (related to the sound
velocity of the system) beyond which the system can lower its energy
by emitting phonon-like excitations which deplete the original
state. This kind of instability, that is closely related to the
energy spectrum of the system, is called \textit{energetic
instability} and it has been observed for a harmonically trapped
condensate in \cite{kett}. On the other hand, due to the interplay
between non-linearity and periodicity in the GP equation governing
the dynamics of the system, for certain values of lattice height and
velocity an arbitrarily small fluctuation of the original state may
grow exponentially in time, thus destroying the initial Bloch state.
This kind of instability, common to many non-linear systems in a
periodic potential, is usually called \textit{dynamical} or
\textit{modulational instability} since it is connected to the
dynamic equation which describes the system. In Fig.~\ref{fig0} we
show a schematic diagram of the stability of the first band of Bloch
states for a condensate in a 1D optical lattice as a function of the
lattice height. The theory predicts the existence of three regions
depending on the the condensate quasimomentum $q$, corresponding to
a regime where Bloch waves are stable solutions, and energetically
or energetically and dynamically unstable.

Although the distinction of these two mechanisms is
straightforward from the theoretical point of view, in the experiments
it is much more difficult to separate energetic and dynamical
instability since, no matter which is the mechanism
responsible for the onset of instability, the original BEC
superfluidity will be compromised and losses of atoms in the ground
state are expected.
Furthermore, as it is shown in Fig.~\ref{fig0}, the stability of the
system depends on the height of the optical lattice: when the
lattice height is greater than the chemical potential of the BEC,
the so-called \textit{tight binding} regime (lattice height $>5\,
E_{rec}$ in Fig.~\ref{fig0}), energetic instability is only a narrow
boundary between the regions of stability and dynamical instability.
On the other hand, when the lattice is weak compared to the chemical
potential, as in the experiments reported in this paper, there is a
wide range of values of quasimomentum for which the system is
energetically but not dynamically unstable. For these reasons the
experimental technique used in the very first experiments devoted to
the study of this regime \cite{burger,cataliottinjp},  namely the
excitation of dipolar oscillations in a periodic plus harmonic
potential, cannot be used to separate the effects of energetic and
dynamical instability. This happens because, during a single
experimental run, the condensate velocity (and therefore its
quasimomentum) evolves throughout the first band and thus explores
both regimes, as indicated in Fig.~\ref{fig0}. The phenomena
observed in \cite{burger} were attributed to energetic instability
because the 1D Gross-Pitaevskii theory used to analyze the data
could not reproduce some of the observed experimental features. This
in turn suggested that finite temperature effects could play a role.
This interpretation lead to a debate \cite{burger} and finally in
\cite{mmod}, by making a comparison with the full 3D theory, it was
shown that the density profiles observed in \cite{burger} could be
attributed to dynamical instability.

In the present paper we show that the precise control of the BEC
quasimomentum, obtained using an optical lattice moving at constant
velocity, is crucial to distinguish the different unstable regimes.
The experimental procedure implemented in this work allows a
comparison with  the predictions of the theory and a study of the
stability of the condensate also in the excited bands of the
periodic potential \cite{di}. The structure of the paper is the
following: in Sec.~\ref{exp} we describe the experimental set-up, in
Sec.~\ref{EI} we present novel results obtained for energetic
instability and discuss the role of thermal excitations, then in
Sec.~\ref{DI} we present a comprehensive investigation of the
phenomenon of dynamical instability and show that it is possible to
clearly distinguish it from energetic instability. In the Appendix
\ref{app} we discuss the effect of the harmonic trapping potential
on the dynamics of the BEC adiabatically loaded in the moving
optical lattice.

\section{Experimental set-up and procedure}\label{exp}
We produce a BEC of $^{87}$Rb  atoms in the hyperfine state $\left|
F=1; m_F=-1 \right. \rangle$ using a standard double MOT apparatus.
Our condensate typically contains $3\times 10^5$~atoms and it is
produced by radiofrequency-induced evaporation of the atomic sample
confined in an elongated magnetic harmonic trap characterized by an
axial and radial frequency of $\omega_z=2\pi \times
(8.74\pm0.03)\,$Hz and $\omega_{\perp}=2\pi \times (85\pm 1)\,$Hz
respectively. The 1D optical lattice is formed by two
counterpropagating laser beams obtained from a Ti:Sa source and it
is aligned along the axis of the magnetic trap. The interference
profile has a spatial period of $\lambda/2$, where $\lambda \simeq
820\,$nm is the wavelength of the laser. Using two single mode
fibers we obtain two gaussian beams with a radius of
$200\,\mathrm{\mu m}$, much larger than the condensate radial size.
In this paper we report measurements carried out with a lattice
height of $s=0.2$ and $s=1.15$, where $s$ is the height of the
optical potential in recoil energies ($E_{rec}=h^2/(2 \lambda^2 m)$,
$m$ being the mass of a Rb atom). The lattice height is calibrated
via Bragg scattering of the condensate and is monitored throughout
the experiment \cite{salomon}. The frequencies of the two beams are
controlled independently by two acousto-optic modulators which use
the same timebase. This allows us to precisely control the velocity
of the lattice $v_{L}$, which is related to the frequency difference
between the two beams, $\Delta \nu$, by the following relation \beq
v_{L} = \frac{\lambda}{2}\,\Delta \nu. \eeq In the laboratory frame
the lattice potential can be written as
\begin{equation}\label{Vlat}
V_{L}(z)=s\, E_{rec} \cos^2\left(\frac{2\pi
  (z-v_{L}t)}{\lambda}\right)
\end{equation}
while the confining magnetic potential is
\begin{equation}\label{Vho}
V_{ho}(\mathbf{x})=\frac{1}{2} m \left( \omega_{\perp}^2
r^2+\omega_z^2 z^2 \right) .
\end{equation}

After producing the condensate in the harmonic potential, we load it
adiabatically into a single Bloch state ramping the intensity of the
lattice from zero to its final value in a time $t_{R} \simeq 2\,$ms.
During the ramp time the lattice velocity is kept constant and no
acceleration is used. This process can be viewed as a deformation of
the spectrum of the system from a free particle one (i.e parabolic)
to a Bloch one, in which the states are labeled by a band index $n$
and a quasimomentum $q$ (that has the periodicity of the reciprocal
lattice, $2\,q_B=4\pi/\lambda=2\,m\,v_B/\hbar$). If the ramping time
is sufficiently long so that this deformation can be considered
adiabatic, the condensate is transferred to a Bloch state with
quasimomentum $q=|v_{L}|\,m/h$, belonging to the first band if
$|v_{L}|<v_B$, to the second band if $v_B<|v_{L}|<2v_B$ and to the
$n^{\mathit{th}}$ band if $(n-1)\,v_B < |v_{L}| < n\,v_B$
\cite{rolston,lensing} (further details are given in the Appendix).
To observe the effect of non-linearity, we maintain a high density
in the sample keeping the harmonic trapping potential on during the
whole experimental procedure. This is crucial to enter the regime
where the effects of dynamical instability are observable.

In order to study the two mechanisms of instability we perform a
time resolved analysis of the losses in the condensate induced by
the lattice for different values of quasimomentum, band index and
lattice height. The general procedure is the following: after
loading the condensate into a single Bloch state we let it interact
with the moving lattice for a time $t$, then we switch off
adiabatically the lattice, release the atoms from the magnetic trap
and measure the number of atoms remaining in the condensate by
taking an absorption image along the radial direction after $28\,$ms
of expansion. This allows us to reconstruct the evolution of the
number of atoms as a function of time and to measure the lifetime of
the condensate $\tau$ as a function of $q$. Without the optical
lattice the BEC lifetime is limited to $\tau = (23 \pm 3)\,$s by the
heating due to current noise in the coils producing the magnetic
trap. The introduction of a stationary lattice (i.e. a standing
wave) introduces two further heating mechanisms: resonant photon
scattering and lattice vibrations due to mechanical noise on the
optics setting the path of the two beams. Using the Ti:Sa laser at a
wavelength of $820\,$nm ($74\,$THz detuned from the D1 line) makes
heating from resonant photon scattering completely negligible, while
we measured that the second effect becomes dominant when the lattice
height is higher than one recoil energy ($\tau = (17 \pm 2)\,$s with
$s=1.15$). In the case of a moving optical lattice this lifetime can
be strongly reduced if instability mechanisms are activated.

%In the presence of a moving optical lattice also for a
In particular we found that, also for a very small velocity, the
lifetime measurement is critically affected by the presence of a
residual cloud of noncondensed atoms surrounding the BEC. Even a
barely detectable fraction of thermal atoms can significantly
shorten the lifetime of the condensate and thus the temperature of
the system must be taken into account. For this reason, in order to
control the temperature of the sample, the radio frequency used in
evaporative cooling is kept on after the production of the BEC
(\textit{RF-shield}).

\section{The role of the thermal fraction: Energetic
  Instability}\label{EI}
\begin{figure*}[ht]
\centering
\includegraphics[width=0.9\textwidth]{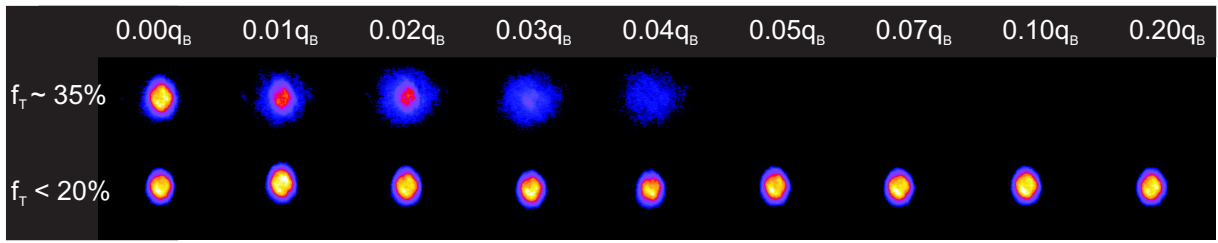}
\caption{Absorption images of the condensate interacting for $t=15$s
  with a lattice with $s=0.2$ for different values of quasimomentum
  ranging from $0$ to $0.20\,q_B$ and for respectively a condensed
  fraction of about $65\%$ (\textit{top}) and no detectable thermal
  component (\textit{bottom}).}\label{fig2}
\end{figure*}

The effect of a thermal fraction can be seen in Fig.~\ref{fig2}.
There we show two series of pictures taken at different lattice
quasimomenta ranging from $0$ to $0.2\,q_B$ and for two values of
the final radiofrequency used for the evaporation, corresponding to
different temperature of the atomic cloud. When we have an almost
pure condensate (thermal fraction $< 20\%$, limited by our imaging
sensitivity), the number of atoms remaining after $15\,$s of
interaction is not sensitive to the lattice velocity (bottom part of
Fig.~\ref{fig2}), while in case of a mixed cloud (thermal fraction
$\simeq 35 \%$) even a small velocity leads to a strong reduction in
the number of atoms (top part of Fig.~\ref{fig2}). Note that the
velocities presented in these pictures are well below the threshold
for the onset of dynamical instability for the experimental
parameters of Fig.~\ref{fig2} corresponding to $q \simeq 0.5\,q_B$
(see Sec.~\ref{DI}); we will come back to this point at the end of
this section. This strong reduction in the number of atoms,
triggered by the dissipation provided by the thermal cloud, can be
qualitatively interpreted as the effect of energetic instability.

\begin{figure}[ht]
\centering
\includegraphics[width=0.98\columnwidth]{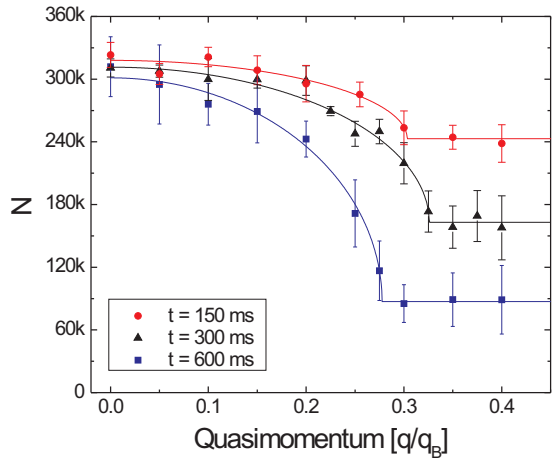}
\caption{Number of atoms remaining in the condensate as a function
of quasimomentum in a lattice with s=0.2 for different interaction
times $t$. The three datasets refer respectively to $t=150\,$ms
(\textit{circles}), $t=300\,$ms (\textit{triangles}) and $t=600\,$ms
(\textit{squares}). The initial cloud has a condensed fraction of
about $65\%$. The curves are fit to experimental points using
Eq.~(\ref{Nt}) with $N_0$, $b$ and $q_0$ as free parameters.
}\label{fig3}
\end{figure}

In order to test this hypothesis quantitatively we have measured the
number of atoms in the condensed fraction as a function of
quasimomentum for three different times of BEC-lattice interaction,
using a $35\%$ thermal fraction. The results are shown in
Fig.~\ref{fig3}. As one can see, the number of atoms $N$ slowly
reduces with increasing $q$ up to a critical quasimomentum $q_0$,
after which it remains constant. The smooth behavior of $N$ as a
function of $q$ below this critical value seems not to be compatible
with the threshold process expected in \cite{wu}. However this
behavior can be understood with an argument similar to the one of
reference \cite{burger}. As we already pointed out, the onset of
energetic instability is related to the sound velocity within the
condensate: once the center of mass velocity exceeds this velocity
the system can lower its energy by emitting phonon-like excitations.
In the presence of axial confinement, as in the experiments reported
here, the system is inhomogeneous and thus the sound velocity, which
depends on the density, is not constant along the direction of the
lattice, vanishing at the edge of the condensate. One can therefore
expect that, once the center of mass velocity $v$ is non-zero, there
exists a fraction of the condensate in which the local sound
velocity $v_s$ is lower than $v$ and therefore that fraction becomes
energetically unstable. The sound velocity for an infinite
cylindrical condensate in a 1D optical lattice is given by
\cite{kraemer} \beq v_s =
\sqrt{\frac{\tilde{g}}{2\,m^*}}\,\sqrt{n_0}, \eeq where $\tilde{g}$
is an effective interaction constant which takes into account the
presence of the lattice, $m^*$ is the effective mass for the
condensate Bloch state and $n_0$ is the peak density of the sample.
Assuming that a similar relation holds for the \textit{local}
density in an inhomogeneous cylindrical condensate, we can argue
that the condensate is locally energetically stable if $v_s(z) >
v_{n,q}$ where $v_{n,q}$ is the Bloch velocity of the condensate
loaded in band $n$ with quasimomentum $q$. As one can verify, for
the values of quasimomentum and lattice height involved in these
measurements, the Bloch dispersion is only slightly different from
the free particle one for which $v_{n,q}=\hbar\,q/m$ and it is
correct to assume $\tilde{g}=g=4\pi\hbar^2a_s/m$ and $m^*=m$ where
$m$ is the mass of a Rubidium atom and $a_s$ is the s-wave
scattering length. The above condition for energetic stability can
be recasted as \beq\label{estab} n_1(z) > \frac{2 \hbar^2 q^2}{g\,m}
= C\, q^2 \eeq where we indicate with $n_1(z)$ the density of the
condensate along its axis. One can therefore obtain the fraction of
the condensate which is energetically stable by integrating the
density over the region satisfying Eq.~(\ref{estab}) \beq\label{fq}
f_{q_0}(q) = \frac{1}{N_0}\int_{n_1>C\,q^2}
\!\!\!\!\!\!\!\!\!dz\,\int\int dx\,dy\, n(\mathbf{r}) \eeq where
$N_0$ is the number of atoms in the condensate and $n(\mathbf{r})$
its density distribution.

In our experiments the lattice produces a weak modification of the
density profile. It is therefore correct to assume as local density
(i.e. averaged over the lattice spacing) the Thomas-Fermi profile of
the condensate in the magnetic potential \beq n(x,y,z) = n_0\,\left(
1 - \frac{x^2 + y^2}{R_{\perp}^2} -\frac{z^2}{R_z^2} \right), \eeq
which gives the fraction of the condensate which is energetically
stable
\begin{eqnarray}\label{eqb}
f_{q_0}(q) & = & \sqrt{1-\left( \qsqz \right)^2}\,\left(1 +
\frac{1}{2}\left( \qsqz \right)^2 + \frac{3}{8} \left( \qsqz
\right)^4 \right) \times \nonumber\\
&\,& \times \theta \left( 1 - \left| \qsqz \right| \right),
\end{eqnarray}
where $\theta(x)$ is the Heaviside function and $q_0 =
\sqrt{(g\,m\, n_0)/(2\,\hbar^2)}$ is the threshold value for
energetic instability in a homogeneous cylindrical condensate with
peak density $n_0$, namely the quasimomentum for which
Eq.~(\ref{estab}) cannot be satisfied for any $z$. The value of
$q_0$ can be theoretically calculated also beyond the
approximation of free particle dispersion and is independent from
the BEC-lattice interaction time provided that the density is not
significantly reduced by the losses induced by the instability.

In order to derive a simple expression for $N$ as a function of $q$
and $t$ one has to make some assumptions on the decay induced by
energetic instability. We assume that for a given $q$ the number of
atoms in the stable fraction (i.e. $N_0\,f_{q_0}(q)$) is constant in
time, while the number of atoms initially in the unstable fraction
(i.e. $N_0\,(1-f_{q_0}(q))$) decays with a time behavior $b(t)$. We
thus obtain for $N$ the following expression \beq\label{Nt} N(q, t)
= N_0\,f_{q_0}(q) + b(t)\,N_0\,(1-f_{q_0}(q)). \eeq Note that
$N_0\,b(t)$ can also be viewed as the number of atoms remaining in
the condensate after a time $t$, once it is entirely unstable
($f_{q_0}=0$) and $N$ no longer depends on $q$. Note also that we do
not make any assumption on the explicit form of $b(t)$, which, for a
given time $t$, enters Eq.~(\ref{Nt}) only as parameter.

\begin{figure}[ht]
\centering
\includegraphics[width=0.95\columnwidth]{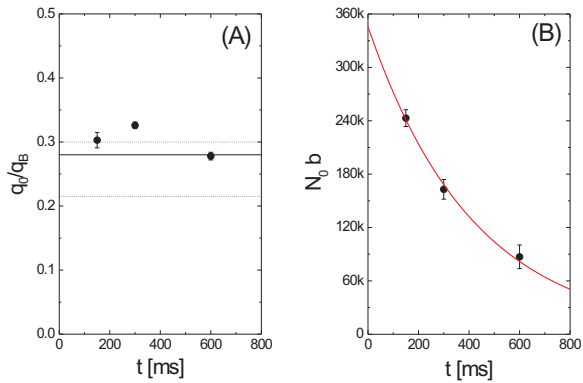}
\caption{(a) Fitted values for $q_0$ at different interaction times
(\textit{symbols}) and theoretical prediction for a homogeneous
cylindrical condensate with the same experimental parameters
(\textit{solid line}). Dashed lines mark the error on the
theoretical value which is due to uncertainty in the measured peak
density of the sample. (b) Number of atoms above $q_0$ given by the
product $N_0b$ obtained from the fit to the experimental data as a
function of interaction time $t$: continuous line is a fit to an
exponential decay from which we measure a characteristic time of
$\tau_{EI}=(416 \pm 55)\,$ms.}\label{figleo}
\end{figure}

The lines shown in Fig.~\ref{fig3} are a fit of Eq.~(\ref{Nt}) to
the experimental data taken for three different values of $t$ with
$N_0$, $b$ and $q_0$ as free parameters. As one can see our simple
model reproduces the experimental points very well within the error
bars which are taken as the standard deviation of a five measurement
average.
In Fig.~\ref{figleo}a we report the values of $q_0$ obtained from
the fits shown in Fig.~\ref{fig3} together with the theoretical
prediction obtained for a homogeneous cylindrical condensate. As
expected, the value of $q_0$ measured for the different times does
not exhibit a significant dependence on the BEC-lattice interaction
time $t$ (see Fig.~\ref{figleo}a). We can directly compare these
values with the theoretical prediction for the threshold of
energetic instability for a homogeneous cylindrical condensate
\cite{mmod}.  Given our uncertainty on the density of the sample,
which propagates into the theoretical calculations, the agreement
between theory and experiment is good. Furthermore one can extract
an estimate for the characteristic time of energetic instability by
looking at the decay of the number of atoms above threshold $b(t)$.
Assuming for $b$ an exponential decay (i.e.
$b=b_0\,e^{-t/\tau_{EI}}$) the timescale for energetic instability
for the experimental condition of Fig.~\ref{fig3} is $\tau_{EI}\sim
400\,$ms, as shown in Fig.~\ref{figleo}b. Due to the experimental
difficulties in controlling the thermal fraction of the initial
atomic cloud, we did not study the dependence of $\tau_{EI}$ on the
condensed fraction. This would be a very interesting measurement
since it could  provide a further insight into the role of the
thermal fraction. However, as we pointed out in the description of
Fig.~\ref{fig2}, as the condensed fraction increases the number of
atoms remaining in the condensate after a fixed $t$ increases as
well.

\begin{figure}[ht]
\centering
\includegraphics[width=0.95\columnwidth]{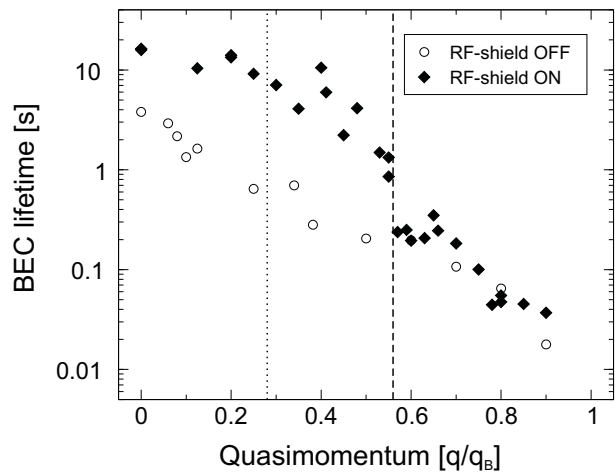}
\caption{Lifetime of the condensate with and without RF-shield as a
  function of quasimomentum for the first Bloch band (logarithmic
  plot). The lattice  height is $s=0.2$ and vertical
  lines are drawn in correspondence of the calculated thresholds for a
  homogeneous cylindrical condensate respectively for energetic
  (\textit{dotted}) and dynamical (\textit{dashed}) instability
  \cite{mmod}. Error bars are smaller than point size.}\label{fig4}
\end{figure}

These results demonstrate that a thermal fraction triggers the
dissipative mechanism connected with energetic instability and
therefore it is possible to strongly reduce this dissipation using a
radiofrequency shield in order to make the measurement with no
discernible thermal fraction. This is particularly important if we
want to separately  address the two regimes of energetic  and
dynamical instability. The results of this procedure are shown in
Fig.~\ref{fig4} where we plot the BEC lifetime as a function of
quasimomentum for $s=0.2$ with and without RF-shield: without the
use of an RF-shield (open circles) it is impossible to distinguish
the onset of dynamical instability at $q=0.56\,q_B$ because this
feature is masked by the reduction of the number of atoms caused by
energetic instability. On the other hand, by keeping an almost pure
condensate throughout the experiment (filled diamonds in
Fig.~\ref{fig4}) the discontinuity in the lifetime entering the
dynamically unstable regime becomes clearly visible despite a
residual reduction of the lifetime due to non complete efficiency of
the RF-shield. An important point is that, even though without
RF-shield there is no measurable threshold, the lifetime deep in the
dynamically unstable regime is the same with or without the
RF-shield as dynamical instability is the dominant loss mechanism in
both cases. We will show in the next section, devoted to a deep
experimental investigation of this regime, that indeed dynamical
instability takes place much faster than energetic instability. We
note here that, as already well established in literature
\cite{corn_rf}, the introduction of an RF-shield increases the
lifetime even in the absence of the optical lattice and this
explains the difference in the measured lifetime at $q=0$.

\section{``Zero temperature'' measurements: Dynamical
  Instability}\label{DI}
For $q>0.5\,q_B$ we observed the onset of dynamical instability
characterized by two distinct signatures which we will discuss in
some detail: a threshold value of quasimomentum above which the
lifetime dramatically decreases (as pointed out above) and, for
higher values of quasimomentum, the presence of complex structures
in the density distribution of the expanded atomic cloud \cite{di}.

\begin{figure}
\centering
\includegraphics[width=0.98\columnwidth]{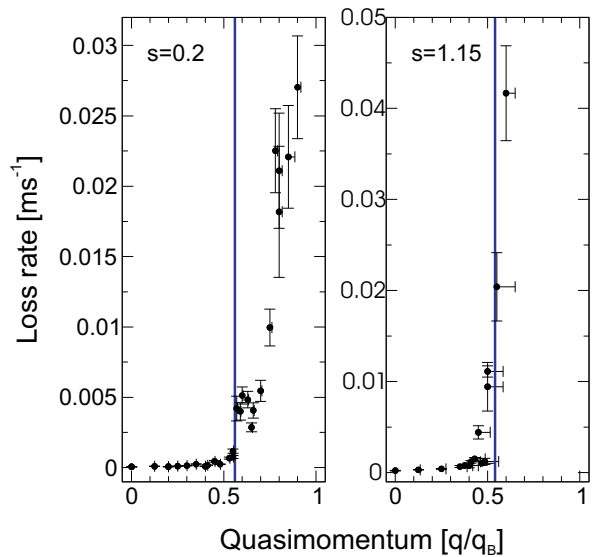}
\caption{Measured loss rate of the BEC (reciprocal of lifetime) as a
  function of quasimomentum in the first Bloch band with $s=0.2$ and
  $s=1.15$ respectively. The vertical lines correspond to the
  theoretical values for the threshold of dynamical
  instability \cite{mmod}. The horizontal error bar is due to the
  presence of the confining harmonic potential as it will be
  explainded in the Appendix.}\label{fig5}
\end{figure}

In Fig.~\ref{fig5} we plot the reciprocal of the lifetime of the
condensate (loss rate) as a function of quasimomentum in the first
Bloch band for two different values of lattice height. For both
values of $s$ we observe a precise value of quasimomentum for which
there is a sudden increase in the loss rate of atoms from the BEC.
For our experimental parameters, this threshold is almost
independent on the condensate density so that, differently from the
case of energetic instability, inhomogeneity does not play a
significant role in this context. Indeed there is a remarkable
agreement between the experiment and the calculated thresholds for a
homogeneous cylindrical condensate shown in Fig. \ref{fig5} as
vertical lines.

\begin{figure}
\centerline{\includegraphics[width=8cm,clip]{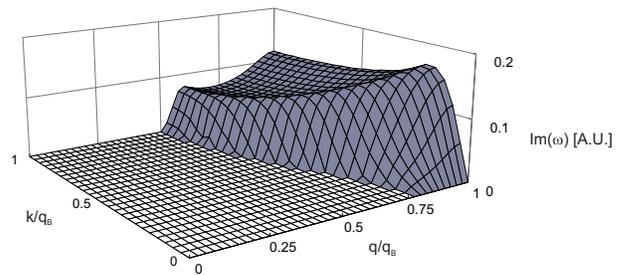}}
\caption{Calculated growth rates $|{\rm Im} (\omega) |$ of the
excitation modes of an infinite cylindrical condensate in a periodic
lattice with $s=1.15$, as a function of  the condensate and
excitation quasimomenta $q$ and $k$ respectively (first Bloch
band).} \label{fig:g_rates}
\end{figure}

From a theoretical point of view the onset of dynamical instability
is signalled by the appearance of excitations of definite wavevector
$k$, that grow exponentially in time, thus modifying the momentum
distribution of the system characterized by peaks spaced by the
periodicity of the lattice in the reciprocal space, $2q_B$
\cite{rolston,pedri}. The character of these excitation modes can be
obtained by considering small deviations from the condensate
wavefunction and solving the corresponding Bogoliubov equations, as
discussed in \cite{wu,menotti} for the 1D case and in \cite{mmod}
for a 3D system. For the system considered in this work when $q
\gtrsim 0.5\,q_B$, the frequency of some modes in the excitation
spectrum develops a nonzero imaginary part $\mathrm{Im}(\omega)$
(see Fig.~\ref{fig:g_rates}), which is the distinctive feature of
dynamical instability and corresponds to the fact that, once
excited, these modes will grow exponentially in time.
As it is shown in Fig.~\ref{fig:g_rates}, for a given value of $q$
there exists a range of such unstable wavevectors $k$, which is
more and more extended as $q$ approaches the band edge. Among all
these unstable modes the one with the highest growth rate plays a
major role in the dynamics of the unstable condensate.

\begin{figure}
\centering
\includegraphics[width=0.9\columnwidth]{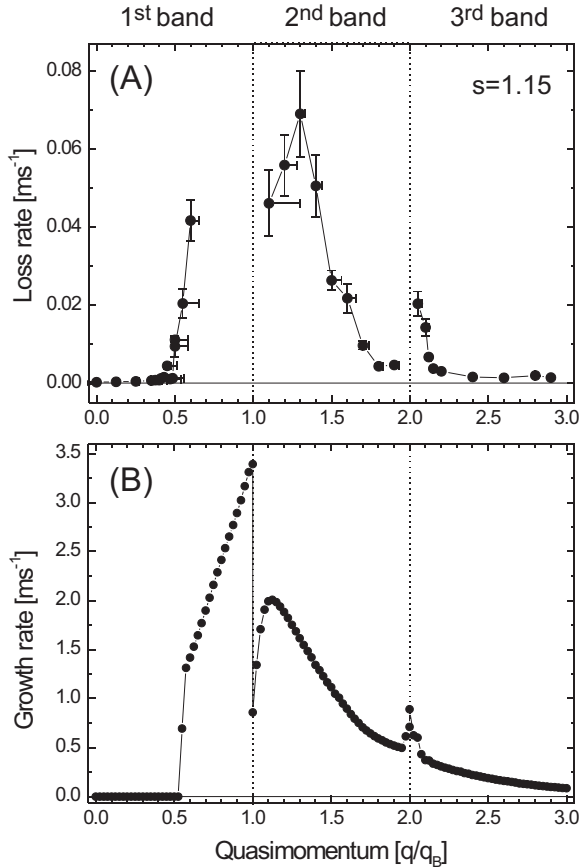}
\caption{Comparison between the measured loss rates of atoms from the
condensate (reciprocal of lifetime) and the theoretically calculated
growth rates of the most unstable modes in the linear regime. Lattice
height is $s=1.15$. Note that the measurements span the first three
Bloch bands.}\label{fig6}
\end{figure}

The unambiguous attribution of the observations to the phenomenon
of dynamical instability has been possible by comparing the
measured loss rates (inverse of the lifetime) to the calculated
growth rates of the most dynamically unstable modes. The results
of this comparison are shown in Fig.~\ref{fig6} for a lattice with
$s=1.15$ . We point out that a full quantitative comparison
between theory and experiment cannot be performed because the
measurements occur outside the validity of the linear analysis on
which the theory is based. However, the distinctive agreement
between the theory and the experiment indeed shows that the
observed phenomenon is dynamical instability and that the mode
which is most unstable in the very initial stage of the dynamics
(i.e. when linear theory is correct) imprints its timescale on the
following dynamics.

\begin{figure}
\centering
\includegraphics[width=0.98\columnwidth,clip]{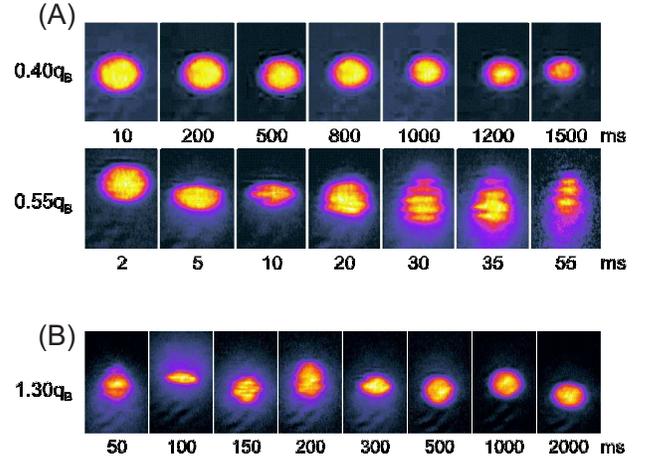}
\caption{(a) Absorption images of the expanded condensate after
different interaction times with a lattice with $s=1.15$ for two
different values of quasimomentum. Note the sudden change of
timescale crossing the threshold of dynamical instability at
$q=0.525\,q_B$ and the appearance of structures in the density
profiles for the unstable case ($q=0.55\,q_B$). (b) Reabsorption of
excitations following $5\,$ms of interaction with the lattice and
different times of evolution in the pure harmonic potential after
switching off of the lattice. In all these pictures the lattice
moves from top to bottom.}\label{fig_profiles}
\end{figure}

\begin{figure}
\centerline{\includegraphics[width=0.95\columnwidth,clip]{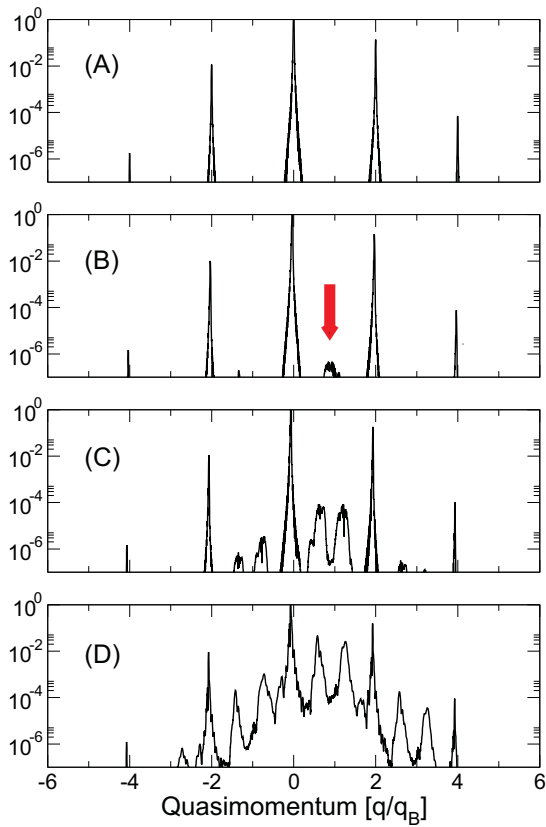}}
\caption{Calculated axial momentum distribution $P(p_z)$ of the
  condensate in the combined harmonic trap plus optical lattice for $t=12, 33, 48, 51$ ms (from
  (a) to (d)) (logarithmic plot). The simulation is performed with
  $s=1.15$. The arrow in (b) marks the appearance of the unstable
  modes.}\label{fig_pspect}
\end{figure}

\begin{figure}
\centering
\includegraphics[width=0.98\columnwidth,clip]{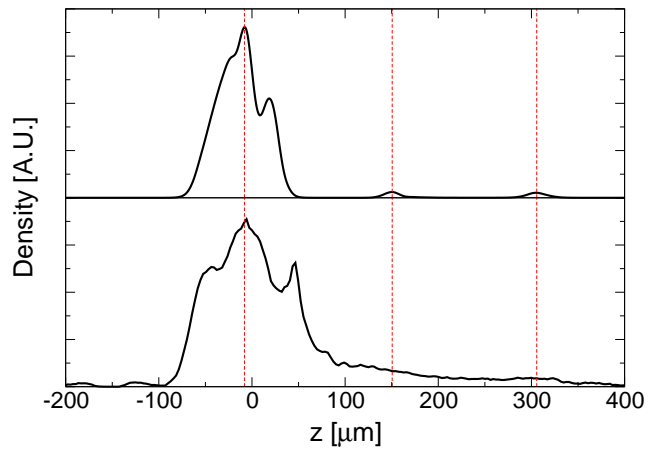}
\caption{Comparison between the simulated
  ballistic expansion of the momentum distribution shown in
  Fig.~\ref{fig_pspect}d (\textit{top}) and the cross-section of the
  measured expanded density distribution of the condensate
  (\textit{bottom}). In both cases the expansion time is
  $28\,$ms, $s=1.15$, $q=0.55\,q_B$, $t=50\,$ms and the lattice is
  adiabatically switched off ramping down its intensity in
  $2\,$ms. Vertical lines mark the three main peaks of the
  calculated momentum distribution corresponding respectively to
  $0$, $1.01$ and $2\,\hbar q_B$. In both cases the lattice is moving in
  the direction of positive $z$.}\label{fig_densdistr}
\end{figure}

For those quasimomenta for which dynamical instability is more
severe and thus lifetimes are shorter, we observed the appearance
of complex structures in the expanded density profiles, shown in
Fig.~\ref{fig_profiles} for $s=1.15$ and $q=0.55\,q_B$.  These can
be the signature either of a density modulation or a phase
fragmentation that leads to the observed fringes through an
interference-like effect \cite{aspect}.
As already stated in \cite{di} we have observed the disappearance of
the fringes as the condensate reverts to its initial state if we let
it evolve in the magnetic potential alone, after switching off the
lattice. This process however takes place in a much longer timescale
($\sim 800\,$ms) than the one of dynamical instability (lifetime can
be of the order of a few milliseconds). In order to compare our
observations with the theory, we have simulated the actual
experimental procedure by solving the time-dependent 3D GP equation
\begin{equation}
\label{eq:gpe}
i\hbar \frac{\partial}{\partial t}\Psi(\mathbf{x},t)=
\biggl[-\frac{\hbar^2}{2m}\nabla^2+V(\mathbf{x},t)+
gN|\Psi|^2\biggr]\Psi(\mathbf{x},t),
\end{equation}
where $V$ is the sum of the harmonic (\ref{Vlat}) and periodic
potential (\ref{Vho}). From the solution of Eq.~(\ref{eq:gpe}) we
extracted the axial power spectrum,  defined as (the tilde
indicates the Fourier transform along z) \cite{nesi}
\begin{equation}
P(p_z)\equiv2\pi\!\int\!\! rdr|\tilde{\Psi}(r,p_z)|^2\,
\end{equation}
which is shown in Fig.~\ref{fig_pspect}.
The upper frame (Fig.~\ref{fig_pspect}a) represents the momentum
distribution at the end of the initial ramp, and is characterized
by sharp peaks localized at integer multiples of $2q_B$ as
discussed above. Then, in accordance with the prediction of the
linear analysys, some modes of complex frequency start growing as
indicated by the arrow in Fig.~\ref{fig_pspect}b (in the numerical
simulation these modes are triggered by the numerical noise).
Afterwards, the nonlinear dynamics introduces processes of mode
mixing and the momentum distribution gets complicated
(Figs.~\ref{fig_pspect}c,d). However the structure is still
characterized  by well localized peaks close to $q_B$, and the
position of the peaks still shows invariance under translation of
$2q_B$.
From these results it is possible to calculate the
expanded density profile and thus make a direct comparison with
the experimental observations. This comparison is shown in
Fig.~\ref{fig_densdistr} for $s=1.15$, $q=0.55\,q_B$,
$t_{R}=10\,$ms and $t=50\,$ms.
The simulation well reproduces the structure of the central peak
observed in the experiment, but the naive expectation that the
expanded density profile of the condensate would simply reflect the
structure of the momentum spectrum is not correct. We note that
indeed the occurrence of dynamical instability may lead to a rapid
population of the noncondensate (thermal) fraction, whose behavior
in the linear regime is governed exactly by the same Bogoliubov
equations as for the semiclassical fluctuations of the condensate
discussed so far \cite{castin97,gardiner}.  This means that the
Gross-Pitaevskii approach may fail when dynamical instability yields
a strong depletion of the ``coherent'' condensed fraction. In order
to account for this, one should include in the theory also the
interaction between the condensate and noncondensate fractions,
which may become macroscopically populated for later times
\cite{stillei}. The formation of a thermal component could strongly
affect the expanded density distribution masking the momentum
component populated by the interaction with the optical lattice.
Actually the experimental density profile reported in
Fig.~\ref{fig_densdistr}(bottom) shows an uniform tail on the right
side of the main peak compatible with a thermal fraction dragged by
the lattice.

\section{Conclusion}
We reported on the experimental observation of energetic and
dynamical instability of a BEC in a moving 1D optical lattice. A
clear separation of the two regimes is obtained by controlling the
temperature of the system and adiabatically loading the condensate
in a Bloch state with precise quasimomentum.

On one hand we have shown that energetic instability is deeply
connected with the presence of a dissipative mechanism such as the
one provided by thermal atoms around the condensate. We have derived
a simple phenomenological model to take into account the effects of
inhomogeneity of the atomic density distribution and found a good
agreement between this model and the experiment. On the other hand
we have observed that in the regime of dynamical instability the
lifetime of the condensate is critically dependent on quasimomentum.
We have compared this distinctive dependence of the condensate loss
rate on quasimomentum with the theoretical prediction on the growth
rate of unstable modes in the initial regime of the dynamics. We
have found a very nice agreement thus demonstrating that the
observed phenomenon is dynamical instability and that the initial
excitations play a dominant role in the following evolution.
Furthermore we compared the structure observed in the expanded
density profile with the results of the solution of time-dependent
3D Gross-Pitaevskii equation finding a significant agreement between
observation and theoretical calculation. We have been able to make a
direct comparison between the timescales of the mechanisms of
instability: we have experimentally demonstrated that in the regime
of low lattice height and for a mostly condensed cloud, dynamical
instability is one order of magnitude faster than energetic
instability. Finally we note that the critical effect of a thermal
cloud on the lifetime of the condensate in the presence of energetic
instability suggests that it is possible to use a moving optical
lattice to detect the presence of a non-condensed fraction well
beyond the sensitivity of the imaging system usually employed in
this field.

\section*{ACKNOWLEDGMENTS}
This work has been supported by the EU under Contracts No.
HPRN-CT-2000-00125, by the INFM Progetto di Ricerca Avanzata
``Photon Matter'' and by the MIUR FIRB 2001. J.~E.~L. was supported
by EU with a Marie Curie Intra-European Fellowship. We thank F.S.
Cataliotti, F. Dalfovo, C. Tozzo and T.W. H\"{a}nsch for stimulating
discussions.

\appendix* \section{The effect of the harmonic potential}\label{app}
\begin{figure}[h!]
\centering
\includegraphics[width=0.95\columnwidth,clip]{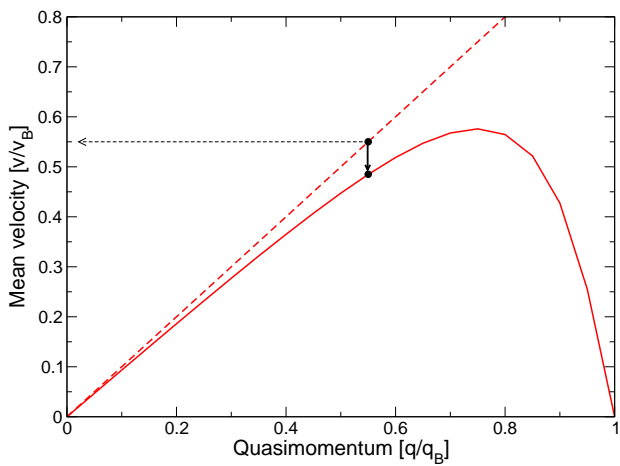}
\caption{Plot of the Bloch velocity as a function of quasimomentum
in the first band for $s=1.15$. The dashed line is the free particle
velocity (i.e. $v=\hbar q/m$). Notice that the deviation of the
Bloch velocity from the free particle one increases as $q$
increases.}\label{fig_ap1}
\end{figure}

\begin{figure}[h!]
\centering
\includegraphics[width=0.95\columnwidth]{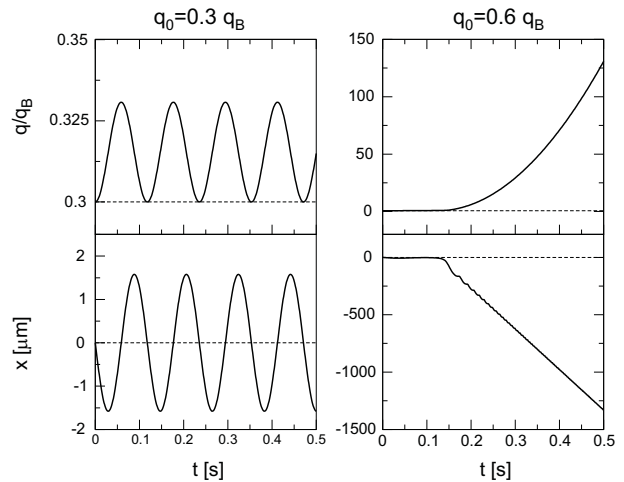}
\caption{Time evolution of a Bloch wavepacket calculated from
semiclassical Eq.~(\ref{eq_ap1}) for $s=1.15$ and two different
quasimomenta $q_0 = 0.3 q_B$ and $q_0 = 0.6 q_B$.}\label{fig_ap2}
\end{figure}

As explained in section \ref{exp}, the adiabatic loading of the
condensate in a moving optical lattice can be viewed as a slow
transformation of the spectrum of the system from the
free-particle parabola to a Bloch band. During this procedure the
wavepacket changes from that of a free-particle to a Bloch state
characterized by a velocity which is related to the energy
spectrum by the well known relation $v_{n,q} =
\hbar^{-1}\,\partial_q E_{n,q}$, where $E_{n,q}$ is the energy of
the eigenstate with quasimomentum $q$ in band $n$. As it is shown
in Fig.~\ref{fig_ap1} the difference between $v_n,q$ and the free
particle velocity increases with increasing $q$.
Since the loading procedure conserves $q$, this means that the
condensate acquires a finite velocity in the laboratory frame and
starts moving in the harmonic potential. The center of mass motion in
the frame of the laboratory follows the semiclassical laws of motion
for a Bloch wavepacket in an external force field \cite{asmer}
\beq\label{eq_ap1} \left\{
  \begin{array}{rcl}
    \dot{z}&=&v_{n,q} - v_{L} = v_{n,q} - \hbar q_0 / m\\
    \hbar \dot{q} &=& F
  \end{array}
\right. \eeq
where $z$ is the direction of the lattice, $m$ is the mass of a Rb
atom, $q_0$ is the initial quasimomentum at which the condensate is
loaded and $F$ is the external force acting on the atoms. In our case
this is the harmonic restoring force $F=-m\omega_z^2 z$. For low
values of $q$, where the band has a parabolic shape (i.e. $E_{n,q} =
\hbar^2 q^2 / 2 m^*$), equations (\ref{eq_ap1}) can be analytically
solved leading to
\beq \left\{
  \begin{array}{rcl}
    z(t)&=& - \sqrt{\frac{m^*}{m^{\phantom{*}}}} \,
      \frac{v_0}{\omega_z} \, \sin \left(
      \sqrt{\frac{m^{\phantom{*}}}{m^*}}\,\omega_z\, t \right)\\
    q(t) &=& q_0 + \frac{m^*\,v_0}{\hbar} \, \left( 1 - \cos\left(
      \sqrt{\frac{m^{\phantom{*}}}{m^*}}\,\omega_z\, t \right) \right)
  \end{array}
\right. \eeq where $v_0 = \hbar q_0 (1/m^* - 1/m)$. This solution is
an oscillation both in real and quasimomentum space at the trap
frequency rescaled by the effective mass $m^*=\hbar\,(\partial_q
v_{n,q})^{-1}$. The amplitude of this oscillation in real space
($\sim 1 \mathrm{\mu m}$) is too small to be measured by our imaging
system while the amplitude of oscillation in momentum space cannot
be neglected. Increasing $q$, when the parabolic approximation of
the band fails Eqs.~(\ref{eq_ap1}) can be still integrated giving
asymmetric oscillations. Eventually, as shown in Fig.~\ref{fig_ap2},
for sufficiently high values of $q$, the solution of
Eqs.~(\ref{eq_ap1}) ceases to be oscillatory: the condensate motion
becomes unbounded and quasimomentum indefinitely grows.
We took into account the motion of the Bloch wavepacket in
quasimomentum space considering an uncertainty on $q$ for data
reported in Figs.~\ref{fig5} and \ref{fig6}, given by the amplitude
of oscillation in $q$ space, as obtained by numerical solution of
Eq.~(\ref{eq_ap1}). These considerations explain also why we did not
take data for values of $q$ close to the band edges for which
dynamics due to the harmonic confinement was not bounded.


\begin{thebibliography}{99}
\bibitem{stillei}
We note that energetic instability, albeit slower than dynamical
instability, is still present and so a thermal fraction is unavoidable
in the experiment.
\end{thebibliography}
\end{document}